# Quantum confinement of Dirac quasiparticles in graphene patterned with subnanometer precision

**Authors:** E. Cortés-del Río [1,2], P. Mallet [4,5], H. González-Herrero [1,2†,] J.L. Lado [6], J. Fernández-Rossier [7,8], J.M. Gómez-Rodríguez [1,2,3], J-Y. Veuillen [4,5] I. Brihuega [1,2,3] *

**Affiliations:**

[1] Departamento Física de la Materia Condensada, Universidad Autónoma de Madrid, E-28049 Madrid, Spain.

[2] Condensed Matter Physics Center (IFIMAC), Universidad Autónoma de Madrid, E-28049 Madrid, Spain.

[3] Instituto Nicolás Cabrera, Universidad Autónoma de Madrid, E-28049 Madrid, Spain

[4] Université Grenoble Alpes, CNRS, Institut Néel, 38000 Grenoble, France.

[5] CNRS, Institut Neel, F-38042 Grenoble, France.

[6] Department of Applied Physics, Aalto University, Espoo, Finland

[7] QuantaLab, International Iberian Nanotechnology Laboratory (INL), Avenida Mestre José Veiga, 4715-310 Braga, Portugal

[8] Departamento de Física Aplicada, Universidad de Alicante, San Vicente del Raspeig 03690, Spain

*Correspondence to: ivan.brihuega@uam.es

Keywords: (graphene, nanopatterning, scanning tunneling microscopy, graphene quantum dots, atomic manipulation)

**Abstract:**
Quantum confinement of graphene Dirac-like electrons in artificially crafted nanometer structures is a long sought goal that would provide a strategy to selectively tune the electronic properties of graphene, including bandgap opening or quantization of energy levels However, creating confining structures with nanometer precision in shape, size and location, remains as an experimental challenge, both for top-down and bottom-up approaches. Moreover, Klein tunneling, offering an escape route to graphene electrons, limits the efficiency of electrostatic confinement. Here, a scanning tunneling microscope (STM) is used to create graphene nanopatterns, with sub-nanometer precision, by the collective manipulation of a large number of H atoms. Individual graphene nanostructures are built at selected locations, with predetermined orientations and shapes, and with dimensions going all the way from 2 nanometers up to 1 micron. The method permits to erase and rebuild the patterns at will, and it can be implemented on different graphene substrates. STM experiments demonstrate that such graphene nanostructures confine very efficiently graphene Dirac quasiparticles, both in zero and one dimensional structures. In graphene quantum dots, perfectly defined energy band gaps up to 0.8 eV are found, that scale as the inverse of the dot's linear dimension, as expected for massless Dirac fermions

**Main Text.**

Graphene is the thinnest material ever realized. It has one dimension reduced to the ultimate one atom limit and the other two extending over a macroscopic scale. Since its discovery in 2004,[1] scientists ambition the controlled reduction of the remaining two dimensions down to the nanoscale, as a powerful route to modify its properties by quantum confining graphene electrons. Early attempts to nanopattern graphene using standard lithography made it possible to confine graphene electrons and to open energy band gaps with the creation graphene nanoribbons and quantum dots[2-5] with very poorly defined edges. Even with today's state of the art technology, it is challenging to improve the patterning resolution below 10nm sizes.[5, 6] Alternative approaches based on growth techniques of islands and ribbons [7-13] do not allow to easily tune the shape of the objects, and usually need a subsequent transfer step when metal substrates are used [11]. The use of STM to nanopattern graphene was also pursued since the early years [14-17] and faint gaps inside graphene nanopatterned structures have been measured [14-16]. Some outstanding advances in this direction have been recently achieved with the creation of circular graphene quantum dots composed of p-n junction rings. [18-22] The generation of a local electrostatic confinement has allowed probing exciting quantum-relativistic graphene properties, such as quasibound states, Berry phase or the development of a "wedding cake" line shape in the density of electronic states. However, the graphene quasiparticle confinement there was limited due to the existence of Klein tunneling [23] and only circular shaped dots with diameter around 100 nm have been studied, the size and shape being imposed by the source of the local potential (a buried charge).

In this work, we have used H atoms as building blocks to define atomically sharp barriers that efficiently confine graphene electrons in all directions. Low energy electronic transport in graphene takes place in the π-orbitals, each carbon atom contributing with one electron. Atomic H chemisorbs on top of graphene carbon atoms, forming a strong s-π covalent bond that effectively removes the π-orbital and one electron of graphene from the conduction band. [24] Large density hydrogenation of graphene is known to open up a very large gap, [25] providing thereby a confinement barrier for massless Dirac fermions. The use of C-H binding to block graphene π-electrons have been envisioned since already a decade, both by removing H atoms from fully hydrogenated graphene [15, 26, 27] and by the patterned adsorption of atomic hydrogen onto a moiré superlattice. [28]

Here, we have developed a new methodology which allows the local manipulation of large ensembles of H atoms (see Supporting Information). First, we perform a macroscopic atomic hydrogenation of the graphene surface by the thermal dissociation of $H_2$. [29] Due to the existence of a physisorption channel, [30] H atoms rapidly diffuse on the graphene surface until they chemisorb, mostly forming highly stable dimers. [30, 31] H dimers correspond to pairs of H atoms adsorbed on neighboring C sites, [30, 31] shown as bright features in STM images of **Figure. 1**a. Next, we stabilize the sample temperature around 220 K. At this temperature, we nanopattern graphene by displacing the STM tip in closed feedback loop, with sample voltage = +4 V and typical tunneling currents of 0.1-1.0 nA, at a constant velocity of 1 nm s$^{-1}$. This creates narrow H patterns along the path followed by the STM tip, as shown in Figures 1.a, b, that show a 22 nm side triangular graphene quantum dot created using this method. The patterning process is schematically illustrated in Figure 1f.

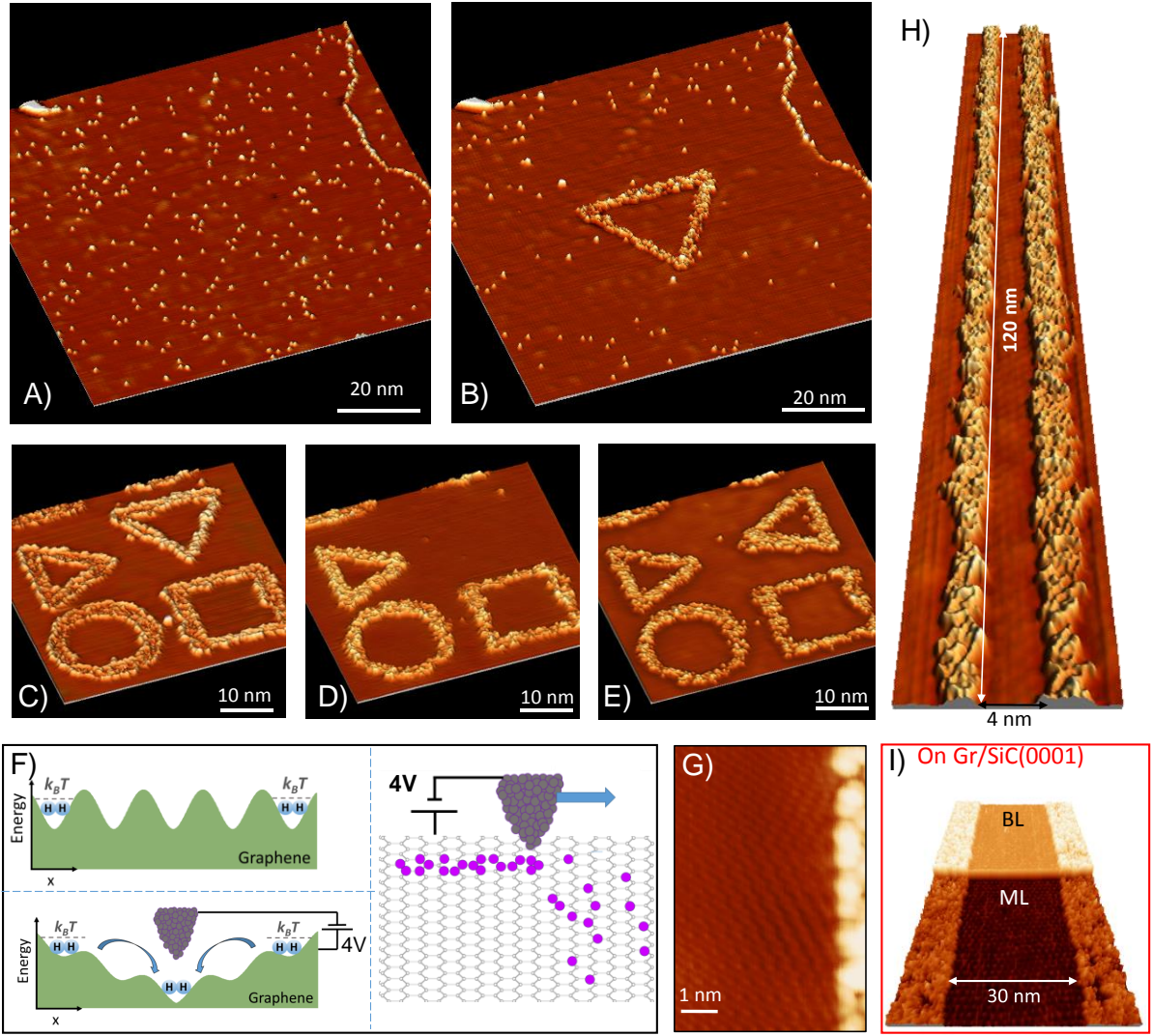

**Figure 1.** Graphene nanopatterning with H atoms. A) 100x100 nm$^2$ STM topography of a graphene region after macroscopic hydrogenation. Isolated bright features correspond to H dimers formed on the graphene terrace. On the right side of the image, H atoms populating a grain boundary serve as a reference. B) Same region as in A) after the creation of a 22nm triangular side graphene dot by H patterning. C-E) Sequence of consecutive STM images where the versatility and reversibility of the patterning method is shown. Nanostructures with different shape and orientation are built-erased-rebuilt in the same 60x60nm$^2$ region. F) Schematic illustration of the H patterning procedure. G) High resolution STM image of a H pattern edge, highlighting the accuracy below 1nm of the patterning process. H) 4nm wide nanoribbon with a length exceeding 100nm. I) 30 nm wide nanoribbon continuously patterned across a terrace step between ML and BL. All STM

images were measured at 220K. Graphene was grown on SiC(000-1), except in I, were it was on SiC(0001).

The working principle of the patterning method is the following. After the adsorption of H atoms (and the subsequent dimer formation aforementioned), setting the sample voltage to +4V seems to modify the energetic landscape at very long distances, favoring the displacement of H towards the tip position. The drift of chemisorbed hydrogen is thermally assisted, as it only works when we rise the sample temperature to ≈220K (see Supporting Information). Figures 1a,b, show two consecutive STM images measured before and after the patterning procedure. It is apparent that H atoms, previously adsorbed on graphene (Figure. 1a), are attracted from very long distances to accumulate along the path followed by the STM tip while operated under patterning tunneling parameters (Figure. 1b). The resulting H patterns are seen by STM as bright protrusions, as in Figure 1b, or as dark depressions depending on the tunneling conditions (see Supporting Information). From our experimental data, we cannot precisely determine the atomic scale arrangement of H atoms inside the H-trenches.

The method is highly robust and versatile. H patterning can be implemented along any arbitrary direction, which enables to build graphene nanostructures with diverse shapes (see in Figure. 1c), such as circles, squares and two triangles with different orientation. Structures with sizes ranging from 1nm up to few μm can be built on demand, just limited by the maximum area the STM can scan. The procedure is completely reversible, in the sense that H patterns can be selectively written-erased-rewritten several times with the same or different shapes. This is illustrated in Figures 1c-e, which shows a sequence of consecutive STM images measured in the same 60x60 $nm^2$ graphene region. First, we patterned 4 different nanostructures, Figure 1c. Next, we removed the upper right triangle, Figure 1d, and finally we patterned, in this same spot, a new triangle with a different orientation, Figure 1e. Notice that the 3 other nanostructures remained almost intact during the whole procedure, despite the harsh tunneling conditions (see Supporting Information) used in the writing/erasing procedure. To selectively remove H patterns we scan the graphene surface with close feedback loop at high speed (≈200 nm $s^{-1}$), low bias (few mV) and high currents (≈1 nA). The H removal procedure is very local, which also enables to partially modify the created structures and, for example, remove a small fraction of the H pattern providing a tunable gateway to graphene electrons.

We have verified that the H patterning methods works on different graphene systems, from multilayer graphene on SiC(000-1) to monolayer (ML) and bilayer (BL) graphene on SiC(0001). This is illustrated in Figure 1i, showing a 30nm wide nanoribbon continuously patterned across a step between a ML and BL graphene grown on SiC(0001). In all these weakly coupled graphene systems, H patterning could be performed by using the same parameters just described. This suggests that it should be equally possible to use our method in graphene over insulating substrates, suitable for electronic gating.

A key point of our method stems in the spatial resolution that we can achieve. As shown in Figures 1g-i, the H patterns that we create are very straight at the nanometer scale. Although we are not yet able to collectively position H atoms down to the last C atom limit, we can comfortably build graphene structures with sub-nanometer precision, as visualized in the zoomed-in images of Figures 1g to 1h . Figure. 1h shows a 4 nm wide nanoribbon with a length exceeding 100 nm and similar length nanoribbons, with 2nm width, can be equally patterned even across a terrace step (see Supporting Information). Close to the patterned H-trenches, (Figure. 1g), graphene shows a ($\sqrt{3} \times \sqrt{3}$)R30º superstructure associated to the existence of intervalley scattering processes, [32] as expected for strong scattering created by sharp barriers.

In the following, we present evidences for an efficient confinement of graphene electrons in such designed structures to highlight the usefulness of this flexible patterning technique for the studies of controlled graphene-based nanostructures. We do not intend to present here a thorough analysis of confinement effects in graphene, but we want to demonstrate that the patterned H structures actually behave as very efficient quantum dots. In order to prove that the H patterns produce quantum confinement of graphene electrons, we have first spatially mapped the electronic structure of graphene inside the created nanostructures. Acquiring conductance maps *dI/dV(x,y)* at different sample voltages *V*, we have directly visualized the quantum confined electronic states inside 1D and 0D nanostructures. In the upper panel of **Figure 2**, we show the results for a 23nm wide nanoribbon fabricated on BL graphene on SiC(0001). We have chosen here BL graphene since, due to pseusdospin, the quasiparticle interferences associated to some scattering processes, in particular those related with intravalley backscattering, are hardly detectable in ML graphene, but are clearly observed in BL graphene [32] (see Supporting Information). This enables the comparison of our experimental results with a simple hard wall model with no free parameters. The model only includes the width of the ribbon, determined with our STM measurements, and the electronic dispersion of graphene electrons in BL graphene on SiC(0001) [32] and it assumes that H patterns act as as impenetrable hard walls for graphene electrons. As it can be seen in Figure 2a, we clearly visualize the first four electronic bound states above the Dirac point inside the ribbon. In the lower panel of Figure 2, we show a 27 nm side square graphene quantum dot. Again, our STS experiments clearly show the existence of several electronic interference patterns, identified as individual quantum confined states by our simple particle in a box model. Despite the relative high temperatures (and hence high thermal broadening) imposed by experimental constrains (170 K), and the simplicity of the model, the close resemblance between experiments and model points to an efficient confinement of graphene electrons inside the created H patterns.

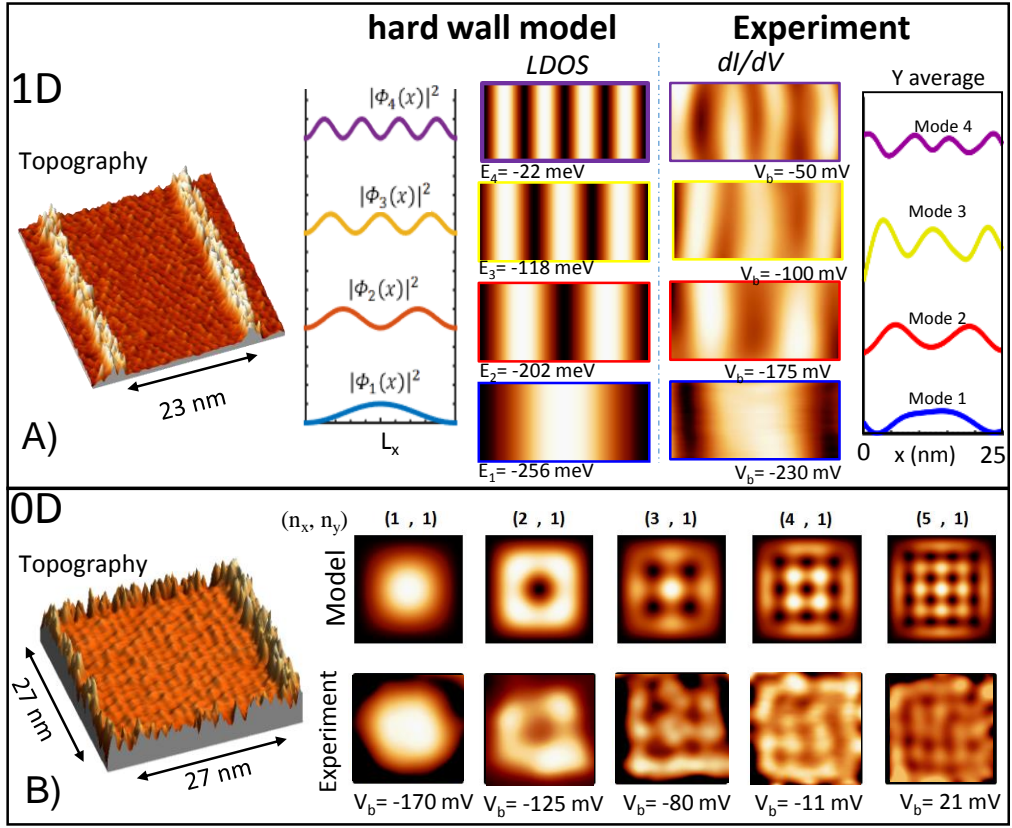

**Figure 2**. Visualization of quantum confined electronic states inside 1D and 0D graphene nanostructures fabricated on BL graphene on SiC(0001). A) STS experiments on a H-patterned graphene nanoribbon. Left, STM image of the 23nm wide ribbon. Right, comparison between experimental conductance images at different sample biases and a simple hard wall model showing the first four quantum confined states. Energies in the model refer to the onset of each confined state. The line profiles show the vertical average of each of the corresponding conductance and LDOS images. B) STS experiments on a patterned graphene nanosquare. Left, STM image of the 27nm wide nanosquare. Right, comparison between experimental conductance images and the simple hard wall model of a particle in a box. $n_x/n_y$ refer to the x/y energy levels.

Ideally, if graphene electrons are perfectly trapped inside a graphene quantum dot, a well-defined energy gap develops and the lifetime of the confined quantum states should be infinite, just limited by thermal effects. To confirm the efficiency of the confinement inside our H patterned graphene quantum dots, we acquired *dI/dV* conductance curves at the lowest achievable sample temperature. We were able to decrease the sample temperature, while maintaining the same scanning region under the tip position, to values as low as 130K, which is a major experimental challenge, considering that the nanostructures have to be created at *T*≈220 K. **Figure 3**a shows a 5nm side triangular graphene quantum dot, patterned on the surface of multilayer graphene grown on SiC(000-1). In this system, the rotational disorder of the graphene layers electronically decouples the π bands, leading to a stacking of essentially isolated graphene sheets, [33-35] with the surface layer equivalent to neutral ML graphene. Reference STS spectra, measured in the clean graphene terrace outside the dot, red curves in Figure 3b, show the characteristic featureless V-shape of neutral graphene. STS curves, measured with the same tip on the center of the triangular dot of Figure 3a, shows the development of a perfectly defined energy gap delimited by two peaks in the conductance, which correspond to the first occupied and empty bound states (see black curves in Figure 3b).

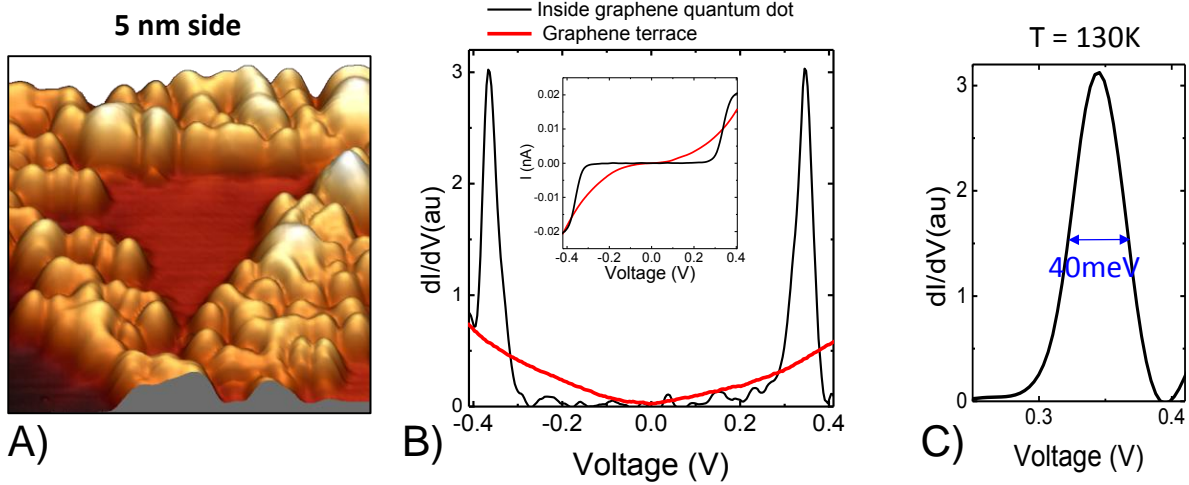

**Figure. 3**. **Strong confinement of graphene electrons inside H-patterned graphene quantum dots.** A) STM image showing a 5nm side triangular graphene dot fabricated on an electronically decoupled graphene layer on SiC(000-1). The graphene surface layer is twisted 15° with respect to the next graphene layer, which ensures electronic decoupling. B) STS curves measured, with the same tip at 130K, in the center of the dot (black curves) and in the outer clean terrace (red curves). A well-defined electronic energy band gap develops inside the triangular dot, proving the strong confinement inside the dot. C) $dI/dV$ curve to show the width of the first empty bound state. The experimental FWHM of 40meV corresponds to the expected thermal broadening at 130K, which implies a very narrow deconvoluted $dI/dV$ peak, associated with a long lifetime of the graphene electron in the bound state.

We would like to stress here that, within experimental resolution, the conductance goes to zero inside the gap. To our knowledge, this is the first STM observation of a patterned graphene quantum dot showing a well-defined gap, which demonstrates that, in our case, graphene electrons are strongly confined in all three directions. An estimation of the lifetime $\tau$ of the confined states inside the dot can also be obtained from STS data. Our experimental data at 130 K shows a FWHM of 40 meV, which can be essentially ascribed to the thermal broadening of STS spectra at such temperature, which is $\approx 3.5 k_B T$ at FWHM. [36] From here, we infer an upper bound for the non-thermal broadening, $\Gamma < 5$ meV for the deconvoluted peaks width. Using $\tau = \hbar\, \Gamma^{-1}$,[37] we obtain a lower limit for the lifetime of the bound states $\tau > 100$ fs. This implies a very long lifetime, and thus a very efficient confinement, for the bound states inside the created dot.

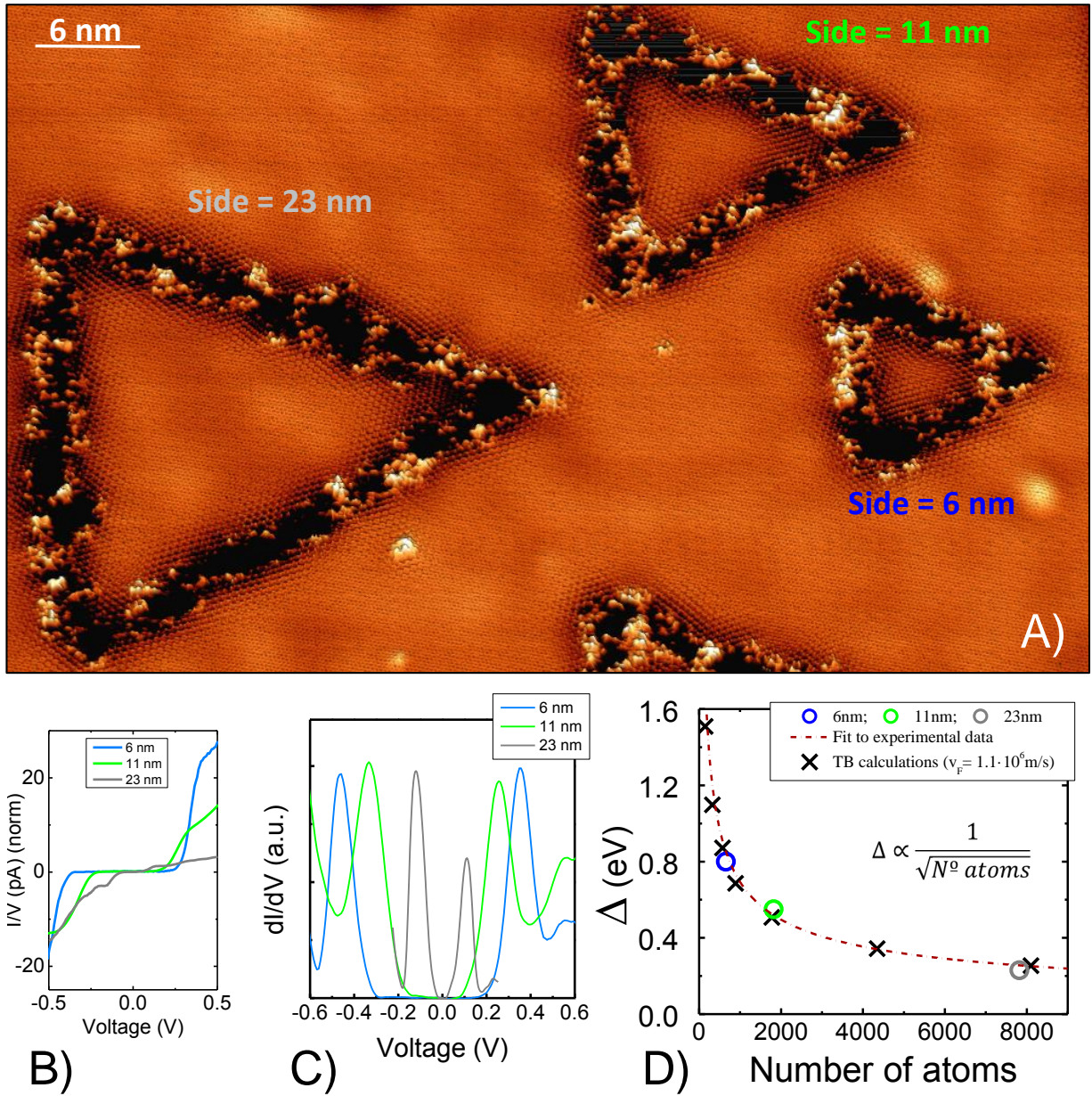

**Figure. 4**. **Tunable graphene energy band gap.** A) 85x60nm² STM image showing three triangular graphene dots patterned, on the same graphene terrace, with different sizes. The sides of the triangles are rotated 11º with respect to the graphene lattice. Triangles are fabricated on a electronically decoupled graphene layer on SiC(000-1), twisted 17º with respect to the next graphene layer. B, C) *I/V* and *dI/dV* curves respectively, measured, at 140K, in the center of each of the triangles shown in A). D) Size dependence of the energy gap Δ, plotted as a function of the number of C atoms inside the dot. Circles correspond to experimental data, crosses to TB simulations and dotted line is a fit to the experimental data.

Early measurements [38] have revealed a size dependent bandgap (scaling approximately as the inverse of the mean sample length) in graphene nanoflakes with uncontrolled shape. Thus, one of the most appealing possibilities offered by the controlled creation of graphene quantum dots, is the opportunity to open tunable energy band gaps in the graphene electronic structure. For equally shaped two-dimensional quantum dots, the gap induced by quantum confinement should scale with the inverse of inverse of $\sqrt{N}$, $N$ being the number of C atoms inside the dots, on account of the linear dispersion of massless Dirac electrons. This scaling law has been verified using ab-initio calculations in triangular graphene nanodots with regular shapes [39]. **Figure 4** shows a terrace of neutral decoupled graphene on SiC(000-1), where 3 graphene triangular dots with sides of 6, 11 and 23 nm have been created. They all have the same orientation with respect to the graphene layer, their sides form 11º with respect to graphene zigzag direction (see Supporting Information). Our STS spectra, measured on the center of the triangles, show the existence of well-defined energy gaps in all three cases, Figures 4b-c. In Figure 4d, we show the evolution of the energy gap $\Delta$ with quantum dot size. We plot the value of $\Delta$, obtained from the energy difference between the maxima of the conductance peaks corresponding to the first occupied and empty bound states, as a function of $N$. A fit to our experimental data, dotted line in Figure 4d, shows that $\Delta$ is proportional to the inverse of $\sqrt{N}$, as expected for massless Dirac quasiparticles.

We have compared our experimental results with an atomistic tight-binding model (see Suppl. Info) that permits to compute the LDOS in triangular graphene quantum dots of different sizes, maintaining the same 11º orientation with respect to the graphene lattice. We have used a value of $t = 3.4\ eV$ for the nearest-neighbor hopping energy, to account for the $v_F = 1.1 \cdot 10^6$ m s$^{-1}$ experimentally found. [40] In line with the experimental approach, we estimate the tight binding value of $\Delta$ from the energy difference between the peak maxima of the first occupied and empty bound states in the LDOS (see Supporting Information). As shown in Figure 4d, there is an excellent agreement between calculated (black crosses) and experimental (open circles) values of $\Delta$, confirming that the patterned nanostructures definitely behave as strongly confining graphene quantum dots (see Supporting Information).

To summarize, we have developed a novel nanopatterning technique that enables to reproducibly build, with subnanometer precision, H trenches acting as hard walls for graphene electrons. As we have demonstrated, this enables to efficiently confine graphene electrons in 1D and 0D graphene nanostructures of complex shapes and the selective opening of energy gaps in the graphene electronic band structure. The versatile tunability of the presented quantum dots and their selective coupling will open a plethora of exciting new possibilities, based on the engineering of artificial Hamiltonians in a graphene platform. In particular, selectively coupled quantum dots can be exploited to engineer artificial topological crystalline phases, higher order topological insulators and frustrated lattices, the latter providing potential playgrounds for exotic correlated states such as quantum spin liquids.

**Experimental Section**
The experimental data here reported were acquired at 220K-130K by using a home-made low temperature scanning tunneling microscope in ultra-high vacuum conditions. Conductance spectra and conductance maps were obtained using a lock-in technique, with an ac voltage (frequency: 830 Hz, amplitude: 5-10 mV rms) added to the dc sample bias. The data were acquired and processed using the WSxM software. [41]

**Acknowledgements**
This work was supported by AEI and FEDER under projects MAT2016-80907-P and MAT2016-77852-C2-2-R (AEI/FEDER, UE), by the Fundación Ramón Areces, and by the Comunidad de Madrid NMAT2D-CM program under grant S2018/NMT-4511, by the Spanish Ministry of Science and Innovation, through the "María de Maeztu" Programme for Units of Excellence in R&D (CEX2018-000805-M). European Union through the FLAG-ERA program HiMagGraphene project PCIN-2015-030; NºANR-15-GRFL-0004) and the Graphene Flagship program (Grant agreement Nº604391). J.L.L acknowledges financial support from the ETH Fellowship program; JFR acknowledges supported by Fundação para a Ciência e a Tecnologia grants P2020-PTDC/FIS-NAN/3668/2014 and TAPEXPL/NTec/0046/2017

# Supporting Information

**Title**
**Quantum confinement of Dirac quasiparticles in graphene patterned with subnanometer precision**

*E. Cortés-del Río, P. Mallet, H. González-Herrero, J.L. Lado, J. Fernández-Rossier, J.M. Gómez-Rodríguez, J-Y. Veuillen, I. Brihuega*[*]

Correspondence to: ivan.brihuega@uam.es

**This file includes:**

Supporting Information consists in 9 sections, labelled from 1 to 9, devoted to:
-**section 1**: Sample preparation and experimental details.
-**section 2:** Procedure to pattern graphene.
-**section 3**: Procedure to remove graphene patterns.
-**section 4**: H nanopatterns on graphene visualized by STM.
-**section 5:** Intervalley scattering at H trenches. Evidence for sharp barriers.
-**section 6**: 2nm wide patterned nanoribbon.
-**section 7**: Calculation details.
-**section 8**: Quantum confinement in bilayer graphene nanostructures.
-**section 9**: Spectroscopy of confined states in triangular islands.

# 1. Sample preparation and experimental details.

A key point of the present work is the atomistic control of the samples, which was obtained by performing all the preparation procedures and measurements under UHV conditions. During the whole process -imaging pristine graphene sample => depositing H atoms on it => and imaging it back- the sample was maintained in the same UHV system.

**Atomic hydrogen deposition:**

We deposited atomic hydrogen following the procedure of refs [1,2], i.e. by the thermal dissociation of $H_2$ on a home-made hot hydrogen atom beam source, see Fig S1. A molecular $H_2$ beam is passed through a W filament held at 1900K. The pristine graphene substrate is placed 10 cm away from the filament, held at RT during atomic H deposition, and subsequently cooled down to 220K, the temperature at which we perform H patterning. After deposition, it takes ~6 minutes to attach the sample into the cold finger of the STM chamber where the measurements are carried out. $H_2$ pressure is regulated by a leak valve and fixed to $\approx 1 \cdot 10^{-7}$ torr as measured in the preparation chamber. The atomic H coverage was adjusted by changing the deposition times to obtain final coverages $\approx 0.1$ H atoms/nm$^2$ (or equivalently, 0.25%; 100%= 38 atoms/nm$^2$ = $3.8*10^{15}$ atoms/cm$^2$, referred to carbon atoms in graphene layers).

After the H deposition the graphene surface presents several bright features (see Fig. 1a) identified as hydrogen atoms, most in the form of H dimers[3,4]. Control experiments, where the sample was exposed to the same $1 \cdot 10^{-7}$ Torr $H_2$ pressure and deposition times keeping the W filament cold and experiments with the W filament at the same 1900K temperature without molecular $H_2$, did not show any trace of H (no bright features could be observed) or any other atomic adsorbate on the surface.

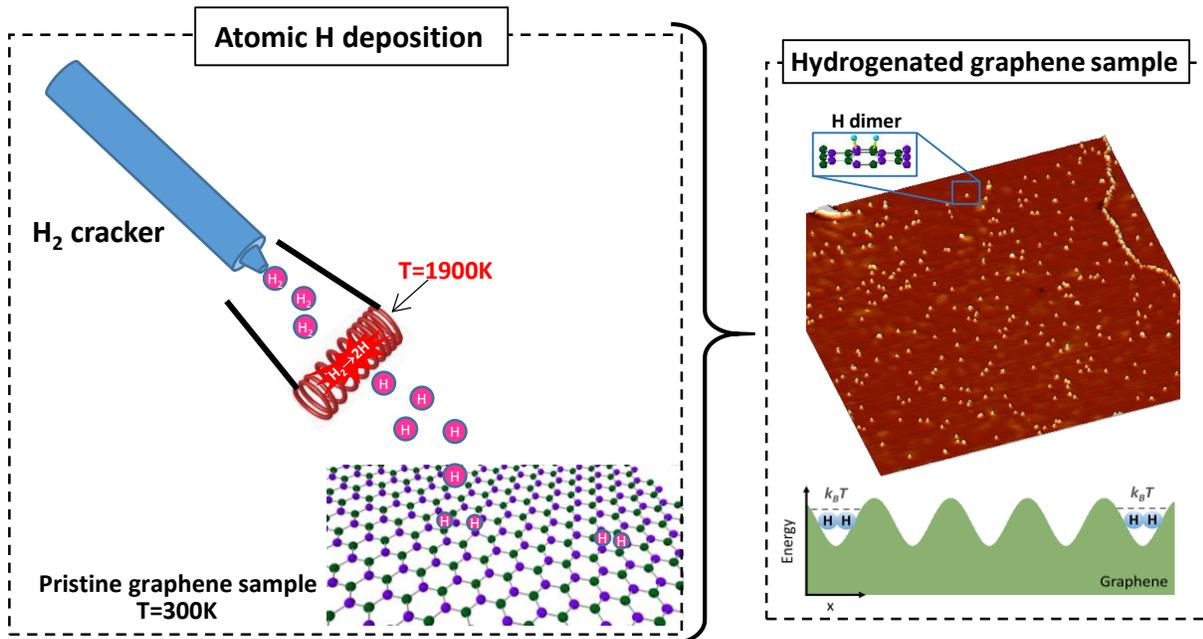

**Fig.** S1. Left: Schematic representation of the H deposition procedure. Right: STM image of a hydrogenated sample, the upper inset correspond to a representative H dimer (see [3,4] for details). The energy diagram at the bottom, illustrates the stability of H dimers, i.e., thermal energy is not enough as to promote H diffusion or desorption.

## 2. Procedure to pattern graphene.

Graphene patterning is a relatively simple and very reproducible process. In the following, we provide a comprehensive recipe with all the experimental parameters needed for H patterning.

- We first deposit H on a clean graphene substrate as detailed in section 1.

- To be able to pattern graphene with H atoms, we need to process the sample at relatively high temperatures to add enough thermal energy to assist the H motion. We typically set the sample temperature to 220K. Temperatures as low as 170K can still be used, although the patterning becomes harder to achieve and the trenches remain uncomplete or open in some parts. For temperatures below 170K, our patterning procedure does not seem to work.

- To attract H atoms, previously adsorbed on graphene, to graphene regions under the STM tip position, we use tunneling parameters of: $It = 0.7$ nA and $Vb$=4V ($Vb$ applied to the sample), Sample voltages between +4V to +5V can also be used, being +4 V the optimal value. Tunneling currents between 0.05nA to 1 nA do also work for patterning.

- To create graphene nanopatterns, we displace the STM tip, in close feedback loop using the tunneling parameters just described, along a predetermined trajectory, with a constant velocity of 0.5nm/s. In this way, H atoms arrange along the trajectory followed by the STM tip, forming the selected graphene nanostructure. A single trace is enough to obtain dense and straight hydrogen nanopatterns. Speeds up to 5nm/s can be used, although we have found that H trenches are denser when slower velocities are used, being 0.5 nm/s an optimal speed for achieving the straightest and densest trenches.

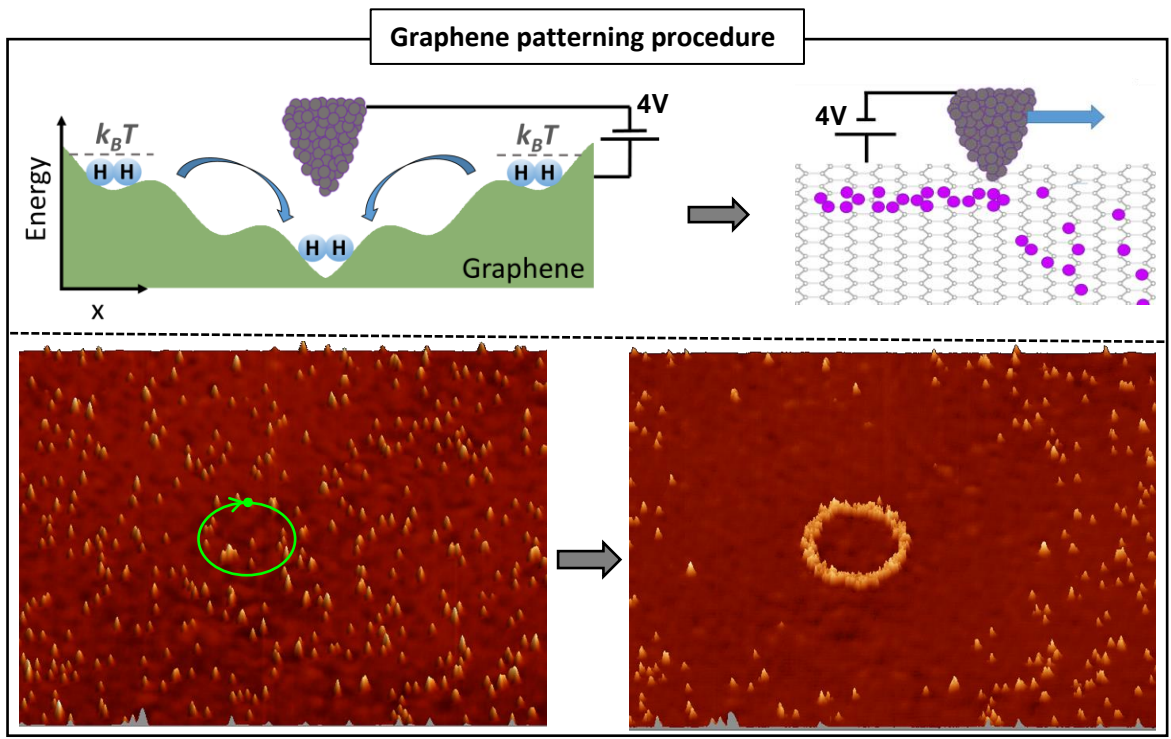

**Fig. S2**. Top: Schematic illustration of the patterning procedure. We promote H motion under the tip position by setting: $It$ =0.7nA and $Vb$= 4V and having enough thermal energy to assists the H motion. With these parameters, we can arrange H atoms along a selected trajectory traced by the STM tip. Bottom: STM images, measured with $It$=0.05nA; $Vb$=1.2V, of the same graphene region before and after creating a 9nm radius circular pattern. Green arrow indicate the trajectory followed by the STM tip, in close feedback loop, with $It = 0.7$ nA and $Vb$=4V.

The STM tips we use are Pt/Ir tips cut with scissors. Our tips systematically get atomic resolution. The nanopatterning procedure worked with 100% of the STM tips that we used. From tip to tip, the "tip manipulation efficiency" showed small variations, which implies that in some cases the tips seemed to be a bit more or less efficient to attract H dimers, probably due to their microscopic shape. In those cases it was enough to slightly adapt the patterning speed to obtain the desired H pattern.

## 3. Procedure to remove graphene patterns.

The created H patterns can be selectively erased by approaching, at low bias, the STM tip towards the sample on top of the H region to be removed. To this end, an efficient approach is to scan across the region of the H trench to be removed, at high speed (100-200 nm/s) in closed feedback loop, using low bias (1-50 mV) and high currents (1 - 10 nA). In this way, it is possible to locally remove H without affecting the surrounding H trench, which enables to partially or totally erase the created structures (see Figs. 1c-e of the main manuscript). This approach can also be used to remove all the H atoms from a selected graphene region (see Fig S3).

This H removal process can be rationalized considering that, according to calculations [1,2], the adsorption energy of a H atom on a Pt tip apex (as our tips are) is around 2.8 eV, which is considerably larger than the adsorption energy for H on graphene (around 1 eV). This indicates that H prefers to adsorb on the Pt tip, the only obstacle being the desorption barrier. This barrier is suppressed by approaching progressively the tip to the deposited H. When retracting the tip, the desorption barrier builds up again, but now with the H adsorbed on the Pt tip [1,2].

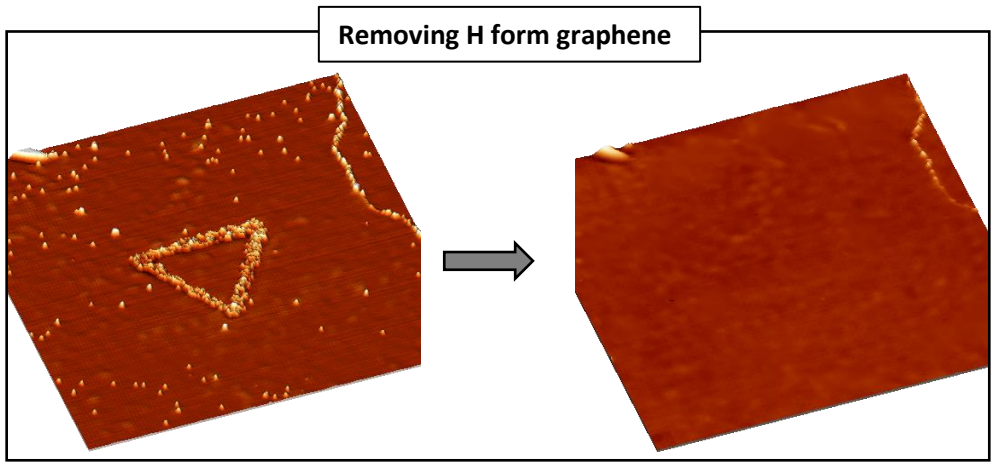

**Fig.S3.** Sequence of consecutive STM images measured on the same graphene region. All H atoms existing in this $60 \times 60 \text{nm}^2$ region, including the triangular nanostructure previously patterned, have been removed by scanning the whole region at 100nm/s with Vb= 3mV and It=9nA.

## 4. H nanopatterns on graphene visualized by STM.

The created H patterns are seen by STM as bright protrusions or as dark depressions depending on the tunneling conditions. Typically, for higher bias voltages (in absolute value) the H trenches are seen as protrusions, while for lower bias voltages they are seen as dark depressions, see Fig S4. However, the observed STM topographic contrast might also depend on the tip apex termination and, in some cases, after an unintentional tip change, an inverted contrast can be appreciated.

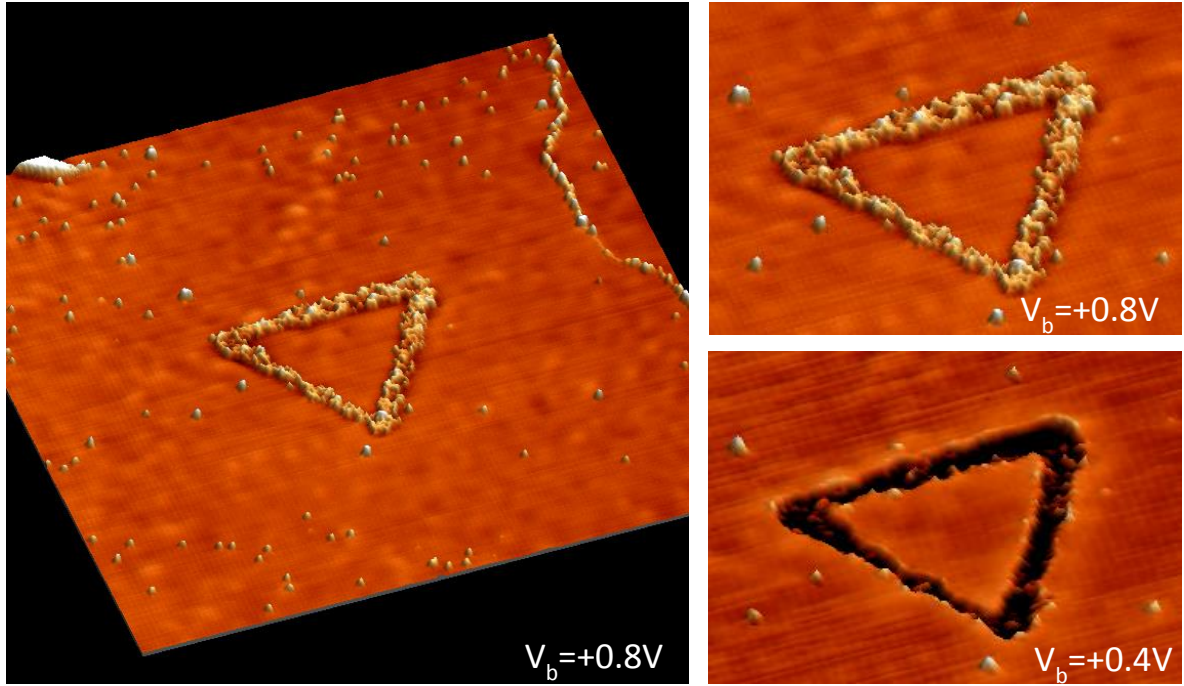

**Fig. S4.** Same region as fig 1b of the main manuscript**.** The created H patterns are seen by STM as bright protrusions or as dark depressions depending on the tunneling conditions. In the right hand side, two STM images of the same triangular structure consecutively measured with the same tip at two different bias values are shown.

## 5. Intervalley scattering at H trenches. Evidence for sharp barriers.

Close to the nanopatterned H-trenches, graphene shows a (√3 ×√3)R30º superstructure associated to the existence of intervalley scattering processes[5], as expected for strongly scattering sharp barriers, see Fig S5 and Fig. 1g of the main manuscript.

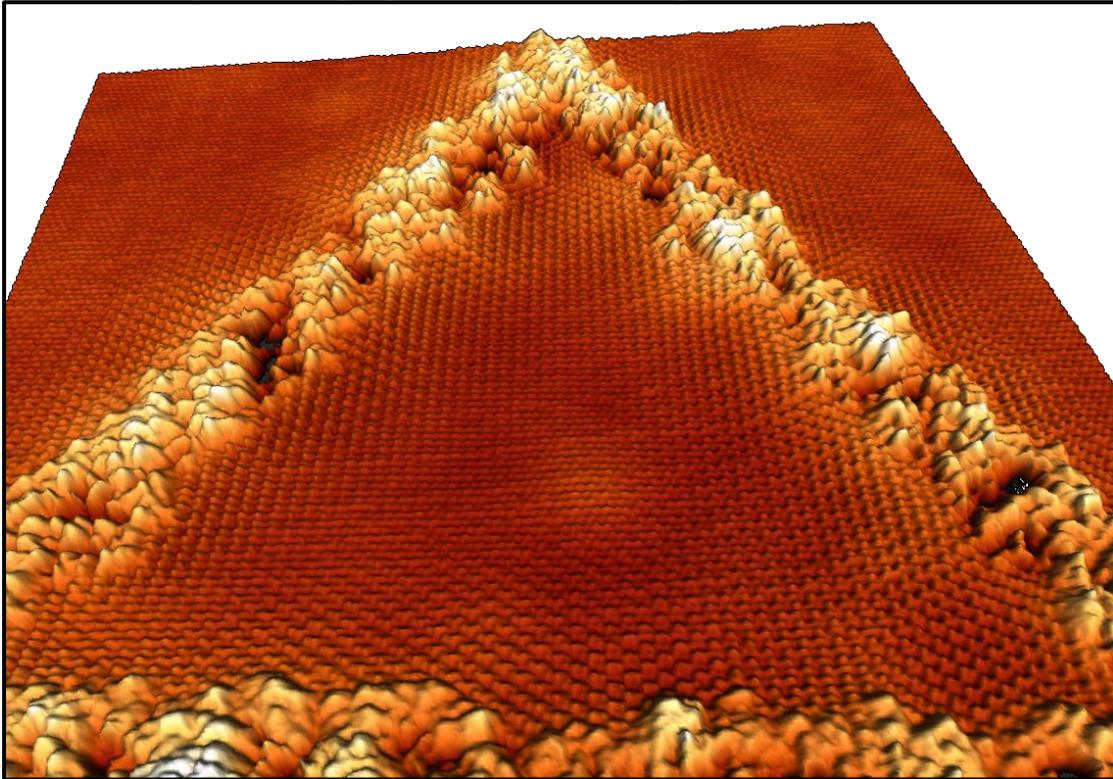

**Fig S5**. STM topography image of a 30nm side triangle with the sides oriented along the zig-zag graphene direction. A (√3 ×√3)R30º periodicity is clearly appreciated emerging form H edges.

## 6. 2nm wide patterned nanoribbon.

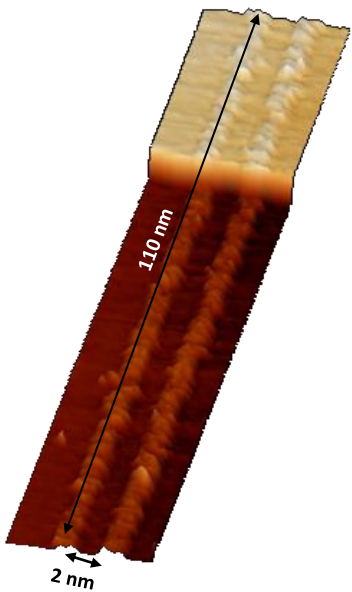

**Fig. S6.** 2nm wide nanoribbon with a length exceeding 100nm patterned across a terrace step.

## 7. Calculation details.

To simulate the experimental data from figure 4 of the main manuscript, we have calculated, as a function of size, the LDOS on the center of graphene triangles with sides forming 11° with respect to graphene zigzag direction. We took a tight binding model with first neighbor hopping of the form

$$H = t \sum_{<ij>} c_i^\dagger c_j$$

where $c_i^\dagger$ creates an electron is site $i$, and $<i,j>$ denotes sum over first neighbors. Given the previous model and a certain geometry of the island, we computed the local density of states defined as $LDOS = <i|\delta(w - H)|i>$, $\delta$ the Dirac delta function.

To reach the largest system sizes, we used the kernel polynomial technique that allows to compute the LDOS for graphene islands with a computational cost linear in the number of atoms the island [6].

In Fig S7, we show the results for a graphene triangle with 1776 carbon atoms. We compute the LDOS value for several C atoms located on the center of the triangle, outlined in red in Fig S7. We have used a value of $t=3.4$eV for the nearest-neighbor hopping energy, to account for the $v_F = 1.1 \cdot 10^6$ m/s experimentally found [7]. Following the experimental approach, we estimate the tight binding value of $\Delta$ from the energy difference between the peak maxima of the first occupied and empty bound states in the LDOS.

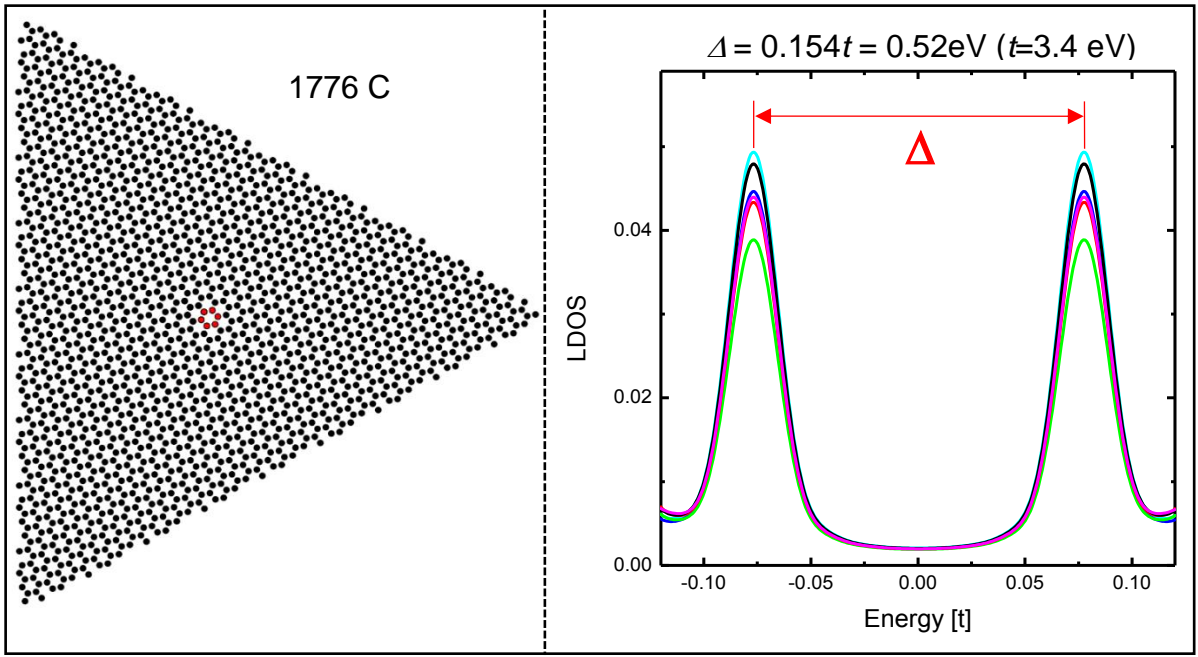

**Fig. S7**. Left, Atomic structure of a triangular graphene island with 1776 C atoms and sides forming 11° with respect to graphene zigzag direction. Right. LDOS computed on the C atoms of the center of the triangle, outlined in red in the left image. The different colors in the graph, correspond to the LDOS of the different C atoms, all of them giving the same Δ value.

## 8. Quantum confinement in bilayer graphene nanostructures.

In the main text, we mention that we have chosen a bilayer graphene (BLG) on SiC(0001) (the SiC Si face) rather than a monolayer graphene (MLG) to reveal the long wavelength oscillations in the local density of states (LDOS) due to intravalley (multiple) scattering at the boundaries of patterned ribbons and squares (Fig. 2 in the MS). Our argument is that due to the peculiar pseudospin in MLG, such oscillations (with period of the order of half the Fermi wavelength $\lambda_F$ at low bias) should be hardly detectable (for MLG on SiC(0001) $\lambda_F \approx 10$ nm (5)). This obviously does not mean that confined modes do not exist in MLG, neither that they do not give rise to any LDOS modulations. Such modes do exist, but they do not lead to long wavelength modulations of the LDOS with significant (sizeable) amplitude. This was demonstrated theoretically for MLG ribbons with either zig-zag or armchair edges for instance in Ref. (8). It was shown (8) that close to the onset of the one dimensional (1D) sub-bands (confined modes), one has either long wavelength LDOS oscillations with *opposite phases* on A and B graphene sublattices (zig-zag edges) or only *short wavelength* oscillations with wavevectors corresponding to intervalley mixing (armchair edges). As a result, the average over a few unit cells perpendicular to the ribbon edges (this is over 1 nm or less) of these LDOS modulations is almost zero for MLG. This is why the LDOS oscillations with period of the order of $\lambda_F/2$ ($\approx 5$ nm at low bias here), which are one fingerprint of the 1D confinement of 2D electrons in free electron gases (9), should not show up in MLG. This behavior is directly related to the pseudospin of the low energy MLG states, which also leads to the disappearance of the long wavelength LDOS modulations at low energy for scattering on point defects (5).

Owing to the coupling between the two Bernal stacked layers, the wavefunctions of the low energy electronic states of BLG are drastically different from those of MLG (5). The pseudospin (as defined in MLG) does not exist in BLG, and accordingly scattering on point defects generate strong long wavelength LDOS modulations with period $\lambda_F/2 \approx 5$ nm at the Fermi level (5). Due to the (apparent) absence of pseudospin effects, we have chosen BLG to illustrate the effect of confinement by H patterns in graphene-based nanostructures, since it resembles the usual behavior observed on regular 2DEG.

The LDOS modulations shown for BLG in the main text correspond to intravalley scattering, this is to the interference of electron states located in the same valley (either at K or K'). At low energy, close to the Fermi level $E_F$, the wavelength of such modulations should be of the order of $\lambda_F/2$, where $\lambda_F$ is the Fermi wavelength of the BLG defined in each K/K' valley. BLG on the SiC(0001) face (Si face) is strongly electron doped, with a Dirac point located about 0.3eV below $E_F$, and with $\lambda_F \approx 10$ nm (see e.g. (5)). To observe several successive confined modes in LDOS maps by STM at (relatively) low bias, the lateral dimension of the nanostructures have to be a few times $\lambda_F/2$. This is why we use ribbons and squares with lateral size of the order of 20-30 nm in figure 2 of the main text.

We restrict ourselves to (relatively) low bias imaging of the BLG on SiC(0001) since it has long been known (see e.g. Ref. (10)) that even at moderate sample bias Vs (i.e. at Vs=-0.5V in (10)) the states from the disordered graphene-SiC interface (the "buffer layer") give a strong contribution to the tunneling current, hindering the detection of the LDOS variations in the BLG. For the same reason, we have concentrated in the "electron like" confined states in the ribbons and squares of Fig. 2, since the the "hole-like" confined states would appear at relatively large negative biases, below the Dirac point (i.e. below Vs=-0.3 V). Notice that in tunneling spectra, the Dirac point of BLG on SiC(0001) appears only as a broad local minimum where the conductance do not reach zero (5, 10), presumably due to this admixture of interface states already for Vs≈-0.3 V.

Finally, we do not present spectra that would reveal the opening of the gap at the Dirac point resulting from the confinement in BLG nanostructures of Fig. 2. The main reason is that a built-in electric field due to the inequivalent doping of the two layers already open a sizeable gap (≈0.1 eV) at the Dirac point is BLG on SiC(0001) (5, 11). Moreover, this layer asymmetry cause a "flattening" of the band dispersion in BLG close to the Dirac point ("Mexican hat") (5, 11). Due to the large lateral size of the nanostructures in Fig. 2, the wavevector $k_0$ of the first confined eigenmode will be small (compared to the Femi wavevector), and because of the "flattened" band dispersion the increase in bandgap size will be small. Consider for instance the ribbon of Fig. 2A with width L=23nm, corresponding to $k_0 \approx \pi/L = 0.014$ Å$^{-1}$. Based on the parameters of our previous work for the band dispersion of BLG (5), we would expect an increase in the band gap by about 12 meV from the confinement in the ribbon (Fig. S8 below). This value is too small to be detected in the present experiment owing to i) the limited resolution due to thermal broadening and ii) the ill-defined shape of the gap quoted above even for the bare BLG.

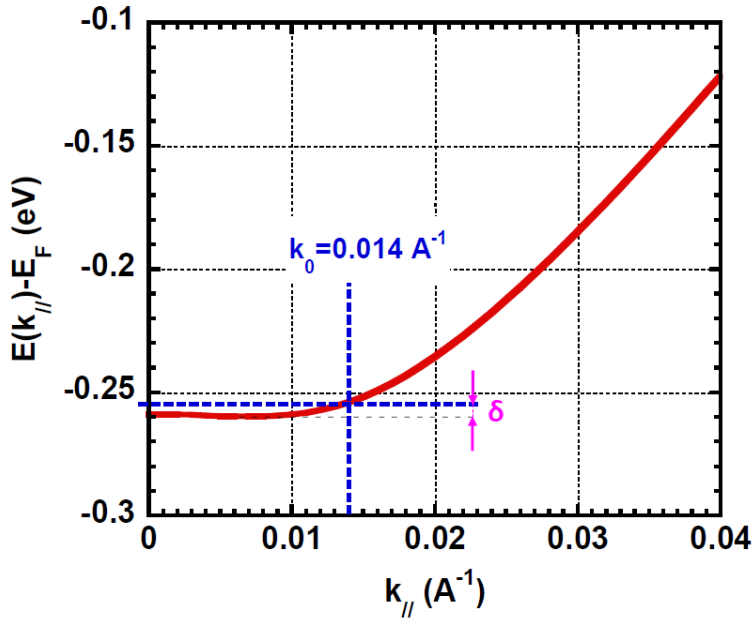

**Fig. S8**. Band dispersion on the lower energy electron-like band of BLG (red line), computed from the parameters of Ref. (5). Here $k_{//}$ is the value of the wavevector counted from the K (or K') point of BLG. The Fermi level is set to 0.0 eV. The Dirac point is at -0.3 eV. The value of $k_0$ estimated for the 23 nm wide ribbon of Fig. 2 leads to a first confined eigenmode located at $\delta \approx 6$ meV above the band minimum. This results in an increase of the bandgap of $2\delta \approx 12$ meV.

# 9. Spectroscopy of confined states in triangular islands.

In figure 4 of the main text, we plot the evolution of the gap of triangular graphene islands with different sizes. We have restricted ourselves to 3 islands with the *same orientation* with respect to the underlying graphene lattice, since this point may be important for establishing the size dependence of the modes. It has indeed been shown in Ref. (12), using tight binding calculations, that for triangular islands with similar sizes the energies of the confined modes was different for armchair and zig-zag edges. In both cases anyway, the energy of all states scale as $1/\sqrt{N}$, where N is the number of atoms in the island, as expected from the linear dispersion of graphene (see main text).

One may wonder whether the peaks observed in Figure 4C could be due to Coulomb blockade. Indeed, we have triangular nanostructures electrically isolated from the surrounding top graphene layer (by the H lines) and capacitively coupled to the STM tip xand to the underlying (twisted) graphene multilayer. This configuration may lead to Coulomb blockade (or dynamical Coulomb blockade) depending on the value of the resistance R between the triangular islands and the substrate, as shown for instance in Ref. (13). The value of R is not known for the case considered here (triangular graphene on graphene), but assuming Coulomb blockade to be present (R>> quantum of resistance), we can perform a rough estimate of Coulomb gap of the system ($\Delta_C$). Considering (13,14) that the tip-island capacitance is smaller than the island-substrate capacitance C, we get $\Delta_C \approx e/C$. C can be estimated from a plane capacitor model for an island of surface S as $C=\varepsilon S/d$, where $\varepsilon$ is the dielectric constant of the graphene-island junction and d is the graphite interlayer distance (d=0.335 nm). A lower bound of C (corresponding to an upper bound for $\Delta_C$) is thus $C=\varepsilon_0 S/d$, with $\varepsilon_0$ the vacuum dielectric constant. The estimated (upper bound) value of the Coulomb blockade gap for a 23 nm side triangular island, as the one shown in figure 4, is $\Delta_C \approx 26$ meV, which almost an order of magnitude smaller than the actual gap value measured in our experiments ($\Delta = 230$ meV). One can thus rule out Coulomb blockade effects as the origin of the peaks measured in our spectra.